\documentclass[a4paper,10pt]{article}
\usepackage{graphicx}
\usepackage{subfigure}
\usepackage{amsmath}
\usepackage{epsfig}
\usepackage{cite}
\usepackage{amsmath}
\usepackage{color}
\usepackage[normalem]{ulem}
\usepackage{authblk}

\title{Atom scattering off a vibrating surface: An example of 
chaotic scattering with three degrees of freedom}

\author[a,b]{Francisco Gonzalez Montoya}
\author[c,d]{Florentino Borondo}
\author[b]{Christof Jung}

\affil[a]{ School of Mathematics, University of Bristol, Bristol, BS8 1UG, United Kingdom }
\affil[b]{ Instituto de Ciencias F\'{i}sicas, Universidad Nacional Aut\'{o}noma de M\'{e}xico, 
Av. Universidad s/n, 62251 Cuernavaca, M\'{e}xico}

\affil[c]{Instituto de Ciencias Matem\'aticas (ICMAT),
Cantoblanco--28049 Madrid, Spain.}

\affil[d]{Departamento de Qu\'\i mica, Universidad Aut\'onoma de Madrid,
Cantoblanco--28049 Madrid, Spain.}

\begin{document}

\date{}
\maketitle

\begin{abstract}

We study the classical chaotic scattering of a He atom off a harmonically vibrating Cu surface. The three degree of freedom (3-dof) model is studied by first considering the non-vibrating 2-dof model for different values of the energy. The set of singularities of the scattering functions shows the structure of the tangle between the stable and unstable manifolds of the fixed point at an infinite distance to the Cu surface in the Poincar\'e map. These invariant manifolds of the 2-dof system and their tangle can be used as a starting point for the construction of the stable and unstable manifolds and their tangle for the 3-dof coupled model. When the surface vibrates, the system has an extra closed degree of freedom and it is possible to represent the 3-dof tangle as deformation of a stack of 2-dof tangles, where the stack parameter is the energy of the 2-dof system. Also for the 3-dof system, the resulting invariant manifolds have the correct dimension to divide the constant total energy manifold. By this construction, it is possible to understand the chaotic scattering phenomena for the 3-dof system from a geometric point of view. We explain the connection between the set of singularities of the scattering function, the Jacobian determinant of the scattering function, the relevant invariant manifolds in the scattering problem, and the cross-section, as well as their behavior when the coupling due to the surface vibration is switched on. In particular, we present in detail the relation between the changes as a function of the energy in the structure of the caustics in the cross-section and the changes in the zero level set of the Jacobian determinant of the scattering function. 

\end{abstract}

\begin{center}
\begin{small}
emails: francisco.glz.mty@gmail.com, f.borondo@uam.es, jung@icf.unam.mx
\end{small}
\end{center}

\section{Introduction} 
  \label{sec:Intro}

The topic of this article is a 3-dof model for the inelastic scattering of a particle offa vibrating surface, a rather complex problem of dynamics. In contrast, the elastic chaotic scattering of an atom by surfaces is a lot simpler and has been studied in detail for Hamiltonian systems with 2-dof \cite{borondo,borondo1,borondo2,guantes}. The 2-dof Hamiltonian systems are easier to investigate because the corresponding Poincar\'e map acts on a domain of dimension 2and is easily represented graphically. In these maps, the stable and unstable manifolds of the most important hyperbolic fixed points built up a homoclinic/heteroclinic tangle, which is the central element of the skeleton of the whole dynamics. The stable and unstable manifolds are of dimension 1 in the domain of the map, i.e., they are also of codimension 1, and as such, they form division lines in the domain which direct the dynamics of the map. Analogous considerations also hold in the corresponding constant energy manifold for the flow since codimensions of stable and unstable manifolds are equal in corresponding maps and flows. For a detailed discussion of the importance of homoclinic/heteroclinic trajectories, see chapter 3 in \cite{wiggins4} and \cite{moser}. A pictorial illustration of the corresponding tangles is presented in chapters 13 and 14 in \cite{abraham}. 

Also for systems with more dof, it is possible to apply analogous geometric and topological ideas to construct dividing surfaces. They are stable and unstable manifolds of higher dimension belonging to invariant subsets also of higher dimension. But they are still of codimension 1 and they built up the corresponding higher dimensional homoclinic/heteroclinic tangles, see the Refs. \cite{kovacs, wiggins3}. There has been recent progress in 3-dof chaotic scattering problems when they are close to a symmetric system, or equivalently, close to a partially integrable system \cite{benet,zapfe,zapfe2,brouzos,gonzalez,Gonzalez_2014}. The essential idea behind this approach is that the principal phase space structures that direct the dynamics in the 3-dof system are quite robust under perturbations of the system. Then the principal structures survive qualitatively and change only slowly under small perturbations of the symmetry. Accordingly, to investigate the nonsymmetric 3-dof system is convenient to start with the symmetric 3-dof system, which can be considered a stack of 2-dof systems, and then turn on smoothly the perturbations, thereby converting the stack of 2-dof systems into the irreducible 3-dof system in which we are really interested.
  
The present article aims to apply this strategy to atom scattering off oscillating surfaces. As a central part of this work, the changes in the caustics in the doubly differential cross-section when the initial particle energy changes are studied. The model studied here is an extension of a 2-dof model proposed in \cite{borondo, borondo1}, and it includes the vibration of the surface as an additional oscillatory dof. Thereby it adds a new closed dof that exchanges energy with the particle and accordingly the chaotic scattering becomes inelastic. 

The important progress of the present work in comparison to previous studies of 3-dof scattering is a detailed step by step analysis of the changes in the topology of regions of continuity of the scattering function as function of the changing initial particle energy and the corresponding changes in the Jacobian determinant of the scattering function and the resulting changes of the rainbow singularities in the doubly differential cross-section.

The organization of the paper is as follows. Section 2 is a brief review of the basic results of the 2-dof model for the elastic collisions of He atoms with a static corrugated Cu-surface. The construction of the scattering functions of the system is presented. Also, the tangle between the stable and unstable manifolds of the fixed point at an infinite distance to the Cu surface in the Poincar\'e map is given for different values of the initial energy of the incident particle. In section 3, the vibrations of the surface are turn on. In subsection 3.1 the scattering function and their Jacobian determinant are studied. Subsection 3.2 explains the construction of the stable and unstable manifolds of the most important invariant subset for the 3-dof system, those dominate the inelastic He scattering off the vibrating Cu surface. In subsection 3.3 the implications for the doubly differential cross-section are presented.
Section 4 contains the final remarks.

\section{The 2-dof model: Elastic He-Cu surface scattering}
  \label{sec:2dof}

The present section contains a study of the basic properties of the 2-dof version of the model previously investigated in detail in \cite{borondo,borondo1,borondo2}. This study serves as preparation for the next section, where the 3-dof system is build up as a pile of 2-dof system.

Let us consider the motion of a He atom coming in from the asymptotic region, and moving towards the interaction region with a static Cu surface. Here and in the following, we use the expression interaction region for the region in position space close to the surface where the interaction between the particle and the surface is important and where a strong deflection of the particle occurs. Let $z$ and $x$ be the spatial coordinates of the atom perpendicular and parallel to the surface, respectively ( the model considers a single tangential coordinate to the surface, i.e. ~in-plane scattering ). The corresponding 2-dof Hamiltonian is
%
\begin{equation}
  H_0(x,z,p_x,p_z) = \frac{{p_x}^2}{2m}+\frac{{p_z}^2}{2m} + V(x,z), 
 \label{eq:1}
\end{equation}
where the function $V(x,z)$ is a corrugated Morse potential with the functional form 
%
\begin{eqnarray}
 V(x,z) & = & D(1 - e^{-\alpha z} )^2 + D e^{-2\alpha z} V_c(x), 
                                 \nonumber \\
 V_c(x) & = & \sum_{n = 1}^{2} r_n \cos{\frac{ 2n\pi x}{a}}. 
 \label{eq:2}
\end{eqnarray}

\begin{center}

\begin{tabular}{l l}

Moser coefficients & $D$ = 6.35 $meV$, $\alpha$ = 1.05 ${\AA}^{-1}$\\

Fourier coefficient & $r_1$ = 0.03, $r_2$ = 0.0004\\

Unitary cell & $a$ = 3.6 $ {\AA}$ 

\end{tabular}

\end{center}

These potential parameters have the same numerical value as the ones in \cite{borondo2} for the Cu surface. Note that the potential is periodic in $x$ direction. Therefore it is possible to treat $x$ as a compact variable restricted to a circle, and in this sense, this dof can be considered a closed one.

In a scattering experiment, the particle starts in the asymptotic region as a free particle and moves towards the interaction region, where the particle remains close to the surface for some time, and finally moves back to the asymptotic region where it again moves freely. The possible complicated behavior of the particle for a finite time and the corresponding sensibility of its trajectory to initial conditions is a phenomenon called transient chaos. For detailed explanations of this phenomenon see \cite{ott, tel,tel2,sanjuan}.

The scattering functions contain essential information to analyze and compare the behavior of different scattering trajectories. These functions map initial asymptotic conditions to final asymptotic quantities, in analogy to the processes in a standard scattering experiment where the initial asymptotes are prepared and the final asymptotes are detected. This natural approach to obtain information on the essential structures in the phase space by a study of trajectories is similar in spirit to other approaches based on trajectories like for example Lagrangian Descriptors and Fast Lyapunov Indicators, see \cite{mancho,mancho1,lega,lega1}. 

The main idea behind this approach to scattering systems is based on the distinction between the trajectories running exactly on the invariant manifolds of localized invariant subsets of the phase space and the general trajectories outside of such particular submanifolds. Generic scattering trajectories come in from the asymptotic region, stay in the interaction region for a finite time only and return to the asymptotic region. In contrast, when a trajectory starts exactly on the stable manifold of a localized invariant subset, then it converges to this subset and does not return to the asymptotic region in a finite time. In this sense, incoming asymptotes belonging to stable manifolds of localized subsets lead to singularities of scattering functions.

This difference allows to identify intersections between these stable manifolds and the domain of the scattering functions as the subset of the domain where the scattering functions are either infinite or are not well defined. Analogous considerations apply to the unstable manifolds by changing the direction of time. Now we work out these considerations in more detail for the present scattering system. 

As a preparatory step, it is necessary to define the appropriate asymptotic variables for the scattering system. In the incoming asymptotic region, a convenient set of initial conditions of a He atom are an impact parameter $b$ measured relative to the $x$ axis and an initial vectorial momentum $\vec{p_0}$ forming an angle $\theta_0$ with respect to the $z$ axis. The corresponding initial energy $E_0$, which is asymptotically the kinetic energy only is fixed by the given initial momentum.
 
\begin{figure}
\begin{center}
\scalebox{0.8}{\includegraphics{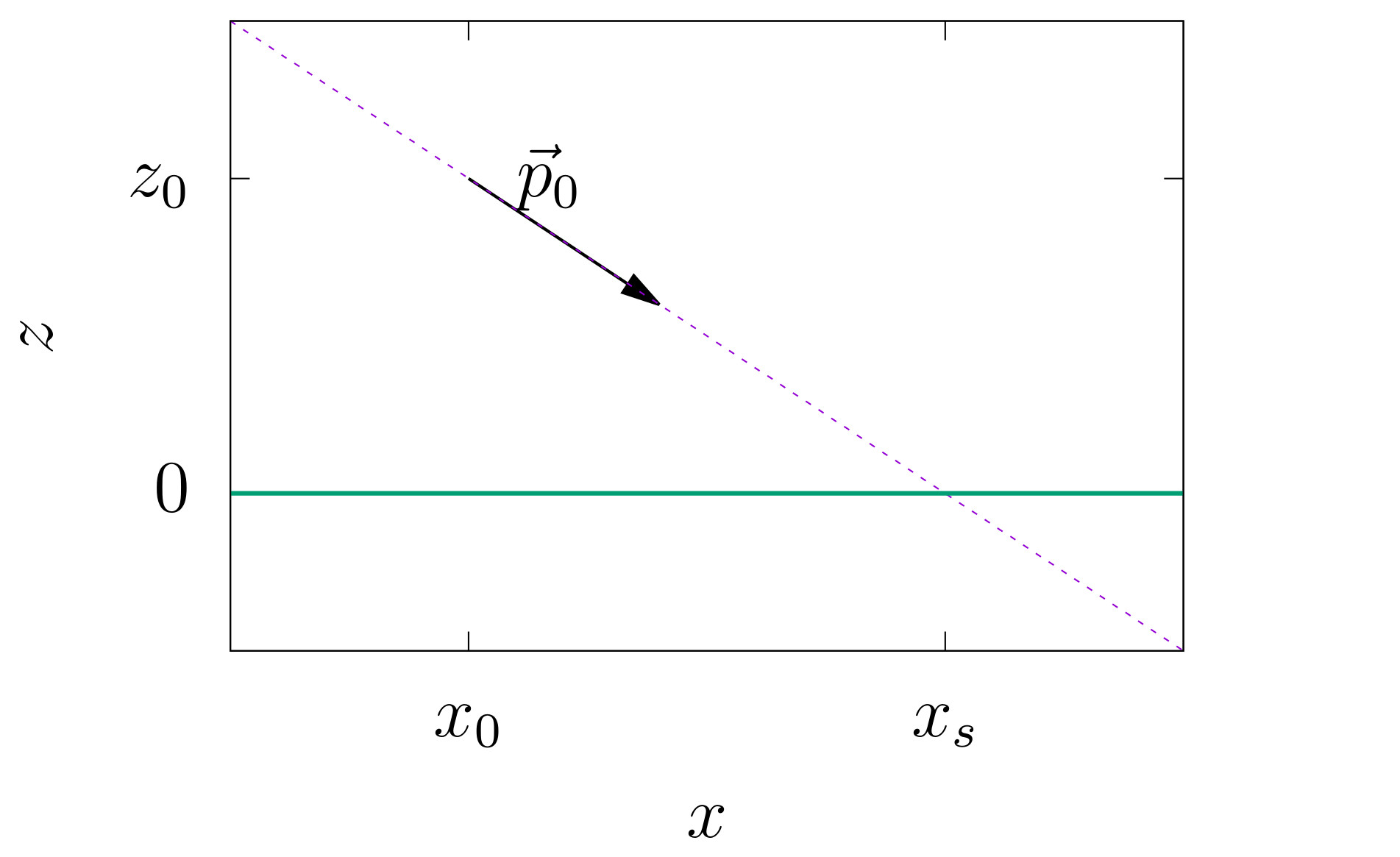}}
\caption{Initial conditions of the scattering trajectories. 
The dotted line shows the intersection for a free particle with the $x$-axis. \label{diagram}}
\end{center}
\end{figure}

In a usual incoming beam, all particles have the same initial energy $E_0$ and the same initial angle $\theta_0$. Equivalently the value of the asymptotic vector momentum is constant over the beam, whereas the initial impact parameter $b$ has a uniform random distribution over the beam. The impact parameter can be defined in the following way: Let us imagine the system without potential. Then the entire trajectory with specific initial conditions would be a straight line. This trajectory would intersect the $x$-axis in a value $x_s$. Since the potential is periodic in $x$, with period $P = a / 2 $, it is only necessary to consider $x$ values $\mod P$. In this spirit, let us define the impact parameter as $b = x_s \mod P$. In addition in a usual beam, the time of arrival at the surface is random. As long as we study the nonvibrating case, this time of arrival is irrelevant. However, it will become essential for the vibrating case considered in the next section. Therefore let us already consider this quantity here carefully. Think of a particle starting with the $z$ value $z_0$ and the value $p_{z,0}$ of $p_z$ at time $t_i$. Then it arrives at $z=0$ at the time $t_s = t_i - m z_0 / p_{z,0}$, where $m$ is the mass of the particle and remember that $p_{z,0}$ is negative for an incoming asymptote. Note that $ t_s - t_i = - m z_0 / p_{z,0} $ is the time of flight from $z=z_0$ to  $z=0$ for a free particle with velocity $ p_z / m $. As consequence we also have $ x_s = x_0 + p_{x,0} ( t_s - t_i ) / m $. These asymptotic variables are illustrated in Fig.~\ref{diagram}. It shows as broken line a free incoming trajectory in the $x$--$z$ plane which passes the point $(x_0, z_0)$ at time $t_i$ and arrives at the point$(x_s,0)$ at time $t_s$. The fat arrow represents the vector momentum $ \vec{p}_0 = ( p_{x,0}, p_{z,0})$ which is conserved along the free trajectory.

The most useful scattering function in the present set up is the final scattering angle $\theta_f$ as a function of the initial impact parameter $b$. 

\begin{figure}
\begin{center}
\scalebox{0.7}{\includegraphics{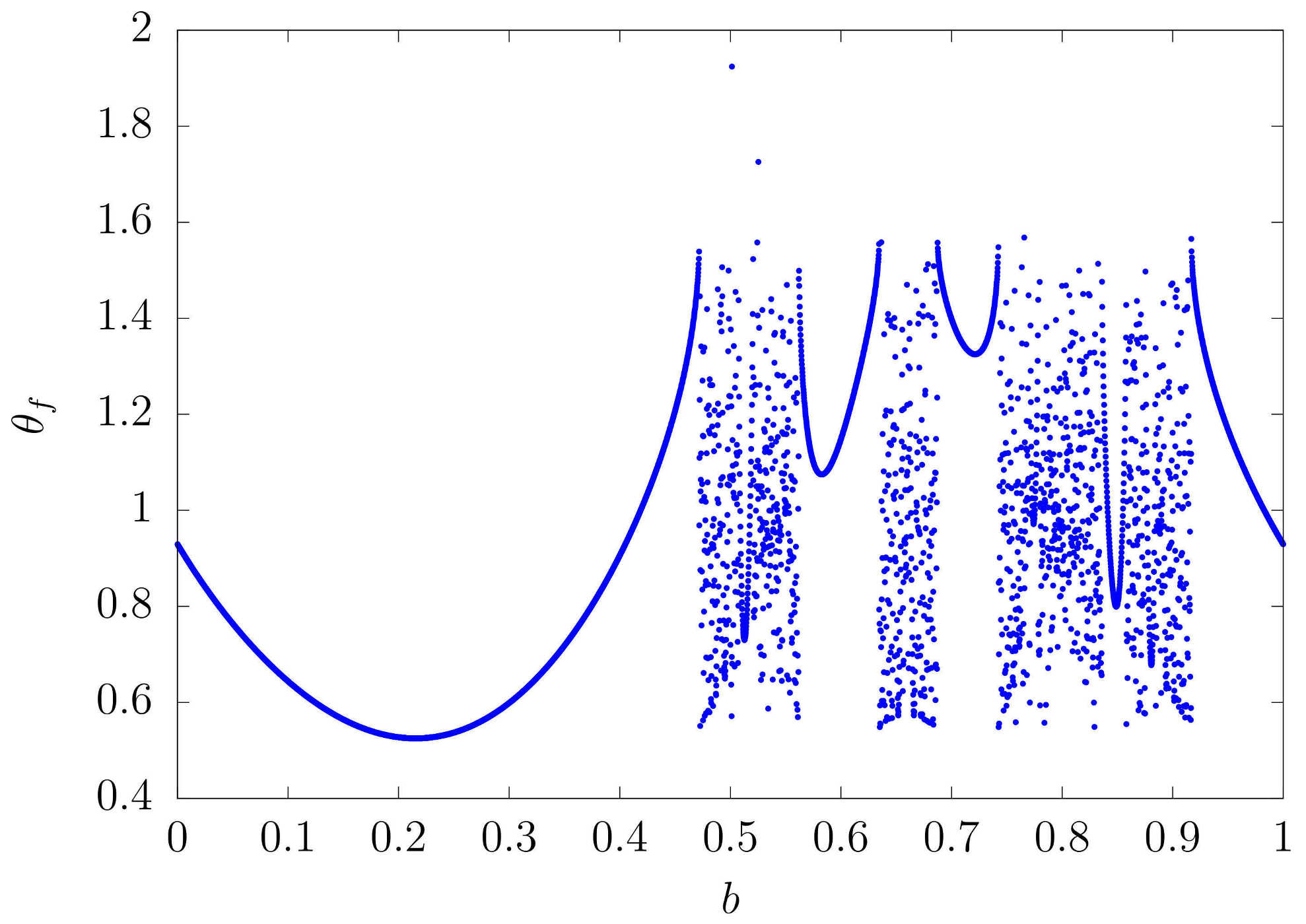}}
\scalebox{1.0}{\includegraphics{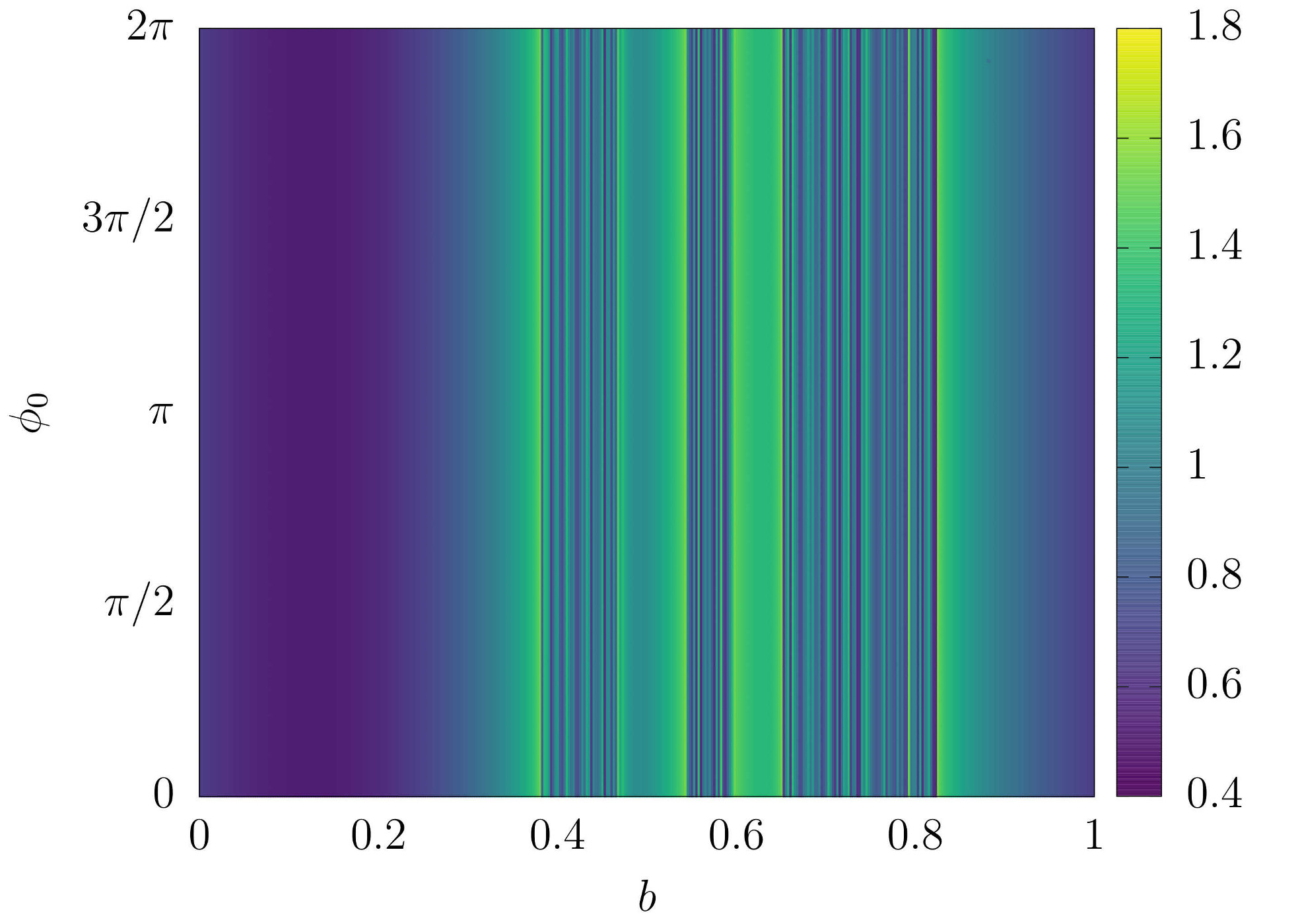}}

\caption{ The scattering functions $\theta_f(b)$ and $\theta_f(\phi_0,b)$ on colour scale for the initial energy $E_0=2$ and the initial angle $\theta_0=80^{\circ}$. The value of the angle $\theta_f$ oscillates as a function of the impact parameter $b$, and it is constant as a function of the phase $\phi_0$ associated with the vibration of the surface, which is taken here as static. The initial phase $\phi_0$ represents the phase shift between the particle motion and the oscillator representing the vibrations, see the section \ref{sec:4dof}. At present, the angle $\phi$ does not influence the scattering process; the surface is static, and the scattering is elastic. Nevertheless, this plot of the scattering function $\theta_f(\phi_0,b)$ will be useful in the next section to compare the elastic and the inelastic system.  \label{S_F_2DF} }

\end{center}
\end{figure}

\begin{figure}
\begin{center}
\scalebox{1}{
\includegraphics[scale=1]{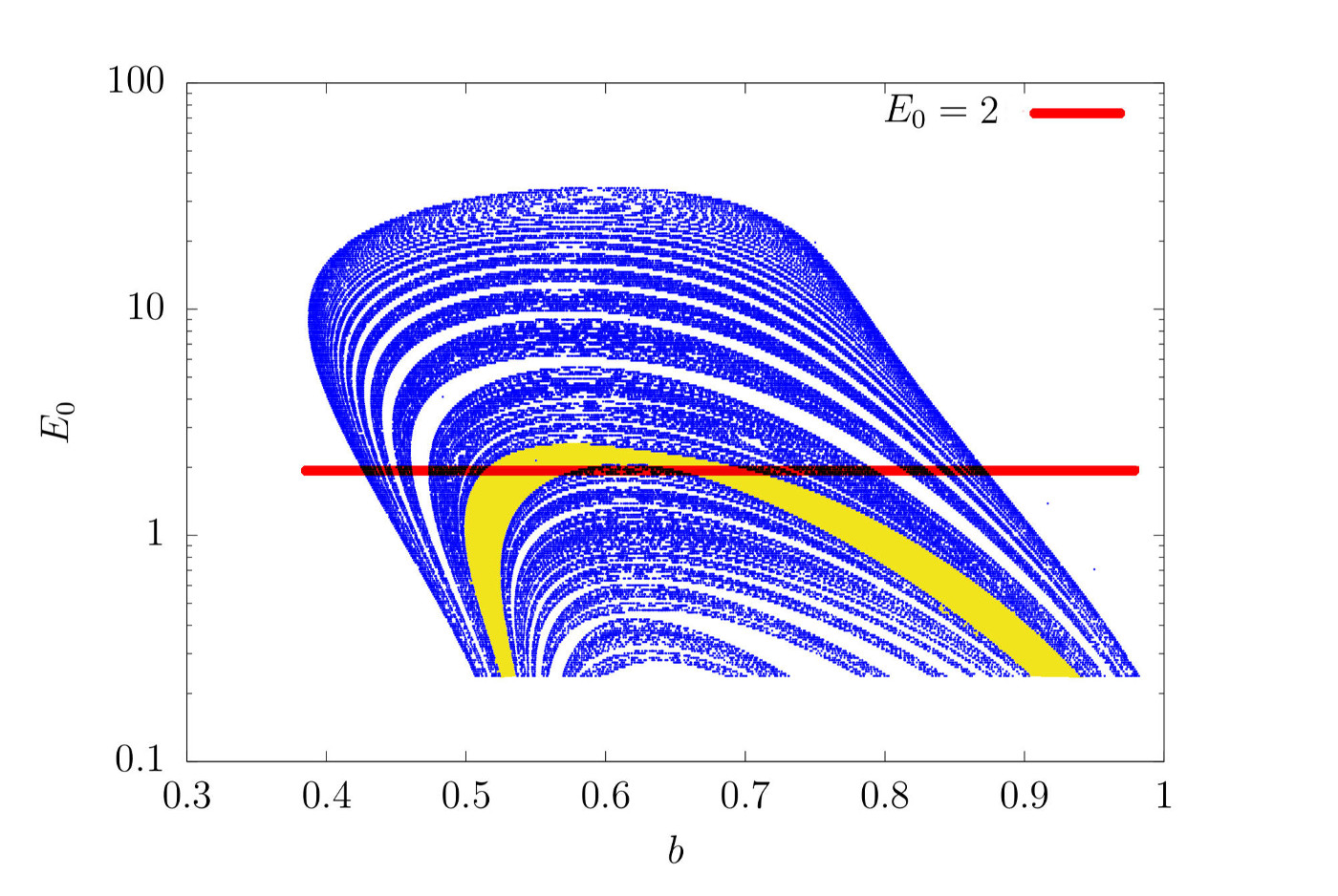}}
\caption{Fractal set of singularities of the scattering function in the $b$--$E_0$ plane, i.e. in the domain of this function for the 2-dof system. We can see the regions of continuity, which are the gaps of the fractal. For every value of the initial energy $E_0$, the singularities reflect the stable manifolds $W^s_{E_0}$ of the chaotic invariant set. The horizontal line at $E_0 = 2$ intersects the same set of singularities as the scattering functions in Fig.~\ref{S_F_2DF}. The region of continuity marked yellow will serve in the subsections 3.1 and 3.3 as example region. 
 \label{F_S_S}} 
\end{center}
\end{figure}

The Fig.~\ref{S_F_2DF} shows the plot of the final scattering angle $\theta_f$ as a function of the initial impact parameter $b_i$ in the upper panel. The initial energy is $ E_0 = 2$ which is a typical value in the energy interval for which the system shows chaos, see Fig.~\ref{F_S_S}, and the initial angle is $\theta_0 = 80^{\circ}$. Along the $b$ axis there is a fractal of points where the scattering function $\theta_f$ is not defined. A point with this property is called a singularity of the scattering function.

In the neighborhood of a singularity, the scattering function shows an infinity of oscillations and an infinity of further singularities. Also, close to a singularity of the scattering function, i.e. in regions where the value of $\theta_f$ changes quickly, trajectories spend more time in the interaction region than trajectories starting far away from the singularities. For a detailed explanation of the scaling behavior of the scattering function and the time delay function in the neighborhood of a singularity, see \cite{jung1}. Furthermore, the $z$-component of the final momentum approaches the value zero when approaching a singularity, equivalently the final angle of inclination $\theta_f$ approaches the value $\pi/2$. 

In Fig.~\ref{F_S_S}, a plot of the position of the fractal set of singularities of $\theta_f$ in the $b$--$E_0$ plane is presented, again the initial angle is $\theta_0=80^{\circ}$. To each point of the set of singularities, there belongs a trajectory that gets trapped forever in the interaction region, and those trajectories lie on the stable manifold of some invariant subset, a ``periodic orbit'' at infinity in this case. First, note the large region of continuity reaching up to infinite values of the energy. In addition, there is an infinite number of regions of continuity for small values of the particle energy $E_0$. In the figure, the largest inner region of continuity is marked by a yellow color, this region serves in subsections 3.1 and 3.3 as an example region of continuity for the case when the perturbation is switched on. The yellow region in Fig.~\ref{F_S_S} corresponds to the two vertical regions of continuity in Fig.~\ref{S_F_2DF}, which are located around the $b$ values 0.6 and 0.72, respectively. In the following, the yellow region and the corresponding regions of continuity in the scattering functions are called the region $R$. In the next section, we will be interested in a variation of $E_0$ such that the number of connected components of the intersection between the red horizontal line and the yellow gap in Fig.~\ref{F_S_S} changes from 2 to 1 to 0. This is the reason why the study is for values of $E_0$ close to 2.

In the general case of a chaotic dynamics, there is a small number of fundamental periodic orbits, such that the tangle built up by the stable and unstable manifolds of these orbits is dense in the whole chaotic invariant set. Then it is sufficient to study the tangle formed by the fundamental orbits only. In most cases, these orbits are the periodic orbits oscillating over the outermost saddles of the potential. However, in the present case, where the potential is of the form given in Eq.~(\ref{eq:2}) there is a small extra problem that the potential has an attractive asymptotic tail, and then the outermost localized orbit of the flow or the outermost fixed point of the corresponding Poincar\'e map sits at infinity.
 
Formally this fixed point at infinity has neutral linear stability, i.e. it is parabolic. However, it is nonlinearly unstable and forms stable and unstable manifolds and tangles (chaotic saddles) of the usual topological structure \cite{zapfe,zapfe2}.

\begin{figure}
 \begin{center}
 
 \includegraphics[scale=1]{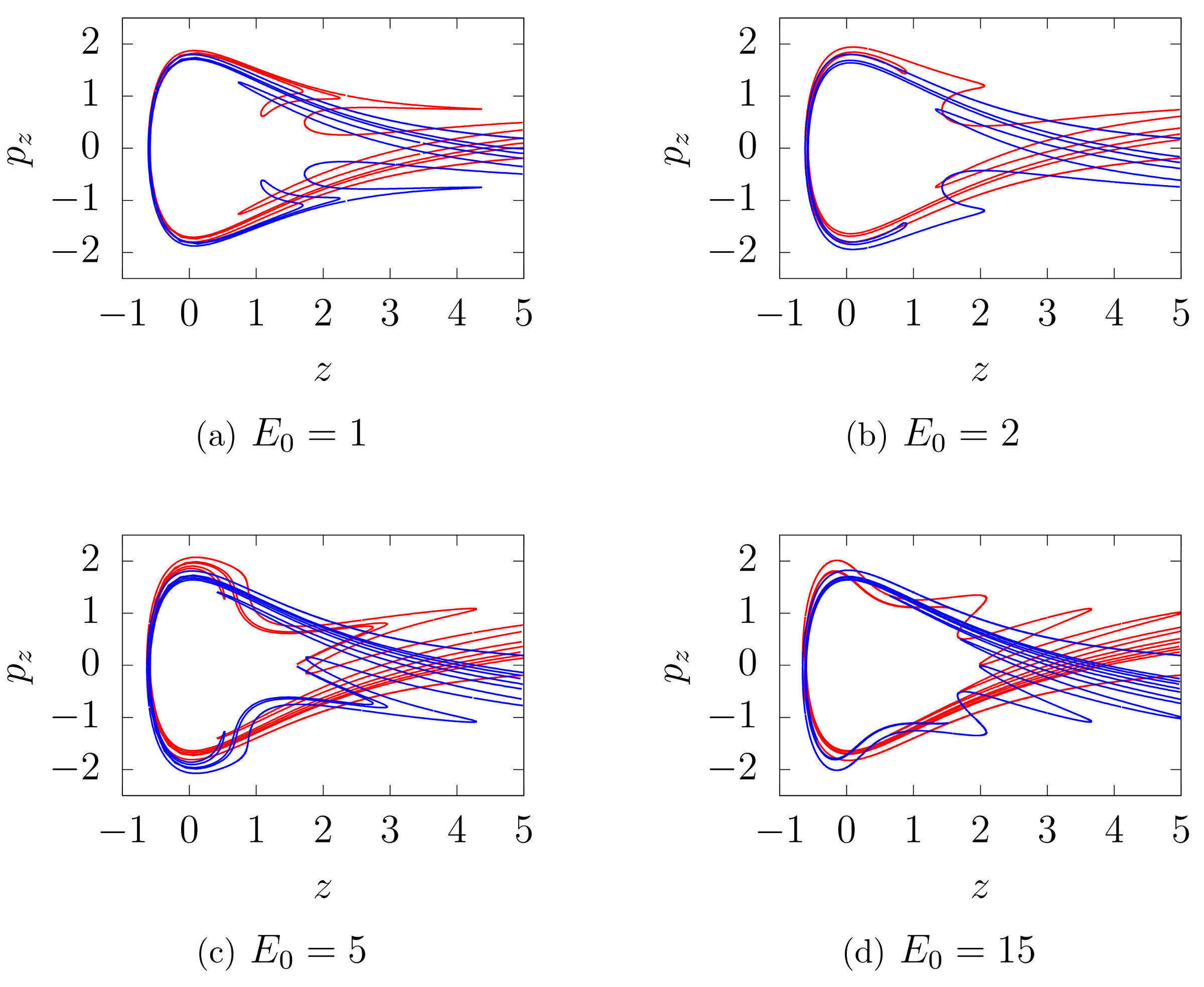}

\caption{ The tangle between the stable and unstable manifolds of the point at infinity for different values of the initial energy $E_0$ which is a conserved quantity for the 2-dof model describing the elastic scattering of He from a static Cu surface. The blue line is a segment of the stable manifold $W^s_{E_0}$, and the red line is a segment of the unstable manifold $W^u_{E_0}$. The tangle between $W^s_{E_0}$ and $W^u_{E_0}$ changes as $E_0$ increases.
\label{H}}
\end{center}
\end{figure}

In an appropriate Poincar\'e section, the unstable manifold is obtained as the reflection of the stable manifold in the line $p_z=0$. This is a consequence of the time-reversal symmetry of the Hamiltonian dynamics. To visualize the tangle, the Poincar\'e map is constructed with the intersection condition $x \mod P =0$. By $W^s_{E_0}$ and $W^u_{E_0}$ are denoted the stable and unstable manifolds of the point at infinity on the Poincar\'e map, i.e the point with $p_z = 0$ and $z \to \infty$. The Fig.~\ref{H} displays the tangle for various values of initial energy $E_0$, and thereby shows the development scenario of this tangle, as a function of $E_0$. The sequence of plots shows the typical scenario of development for a binary Smale horseshoe.

Finally, let us consider the dimension of the important geometrical objects in the phase space relevant to the scattering process. The domain of the Poincar\'e map has dimension 2, the dimension of the stable and unstable manifolds $W^s_{E_0}$ and $W^u_{E_0}$ of the fixed points in the map is 1, i.e. they are of codimension 1. Therefore, these manifolds divide the domain into separate pieces and they are natural barriers that direct and channel the dynamics of the map. The dynamics of points in a lobe delimited by segments of stable and unstable manifolds are determined by the structure of the tangle. The image of a lobe under the Poincar\'e map is the next lobe \cite{meiss}. Moreover, the iterations of the lobes rotate around an inner fixed point in the Poincar\'e map. This fixed point corresponds to a periodic orbit trapped in the periodic potential of the surface \cite{borondo2}. 
 
The situation for the flow in the phase space restricted to a surface of constant energy is equivalent. Here all relevant dimensions are higher by 1, but the codimensions remain the same. The constant energy surface has dimension 3, the stable and unstable manifolds have dimension 2, and they form a partition of the domain. These considerations will turn out useful in the following because the codimensions of the important objects that direct the dynamics stay the same when the vibration of the surface is included, and the system becomes 3-dof.

\section{The 3-dof model: the inelastic scattering case}
  \label{sec:4dof}

In order to construct the 3-dof Hamiltonian associated with the vibrating copper surface, we include an oscillatory term with a single frequency $\omega$. Then, to convert the system into a time-independent one, $\omega t$ is replaced by the phase variable $\psi$ of an oscillator and includes its conjugate action variable $I$ into the Hamiltonian. Once the system is transformed in a time-independent 3-dof version, it is possible to apply the ideas developed in Refs.\cite{zapfe,zapfe2} to study the scattering process.

The construction of the 3-dof model starts from the time-independent 2-dof one. To include the surface oscillation, the variable $z$ in the potential from Eq.2 is replaced by $ z + B_z \cos{\omega t}$. After that,  $\omega t$ is replaced by the phase coordinate $\psi$ of the oscillator and next approximation is taken

\[ V(x,z + B_z \cos{\psi} ) \approx V(x,z) +  B_z\cos{ \psi}\frac{\partial 
V(x,z)}{\partial z} . \]    

Thereby the new potential consists of two terms, first the old potential from Eq.2, and second an additional potential that describes the interaction between the oscillator and the particle. The resulting Hamiltonian model of the complete system is the sum of three terms: The Hamiltonian of the 2-dof model (see Eqs. (1) and (2)), a term $ \omega I $ for the free oscillator, and finally an interaction between the oscillator and the particle.

\begin{eqnarray}
 H(x,z,\psi,p_x,p_z,I) & = & \frac{1}{2m}\left(p_x^2+p_z^2\right)+V(x,z)+ I\omega + 
   \nonumber \\
   & &   B_z \cos{ \psi} \frac{\partial V(x,z)}{\partial z}. 
 \label{eq:3}
\end{eqnarray}

The dof associated with the vibration of the surface is a closed one by construction. The potential $V(x,z)$ is periodic in $x$, and accordingly, the associated $x$-dof can also be considered to be a closed one. In this sense, the 3-dof system has 1 open and 2 closed dofs. The main effect of the interaction term is an energy transfer between the surface oscillator and the particle. 
Therefore, the final particle energy is, in general, no longer equal to the initial particle energy. Fora simple model to explain the energy transfer between a vibrating target and a scattered particle, see Refs. \cite{papa1,papa2}.

In all scattering problems, it is important to make an appropriate choice of the asymptotic labels. For an $n$-dof Hamiltonian system, the asymptotes should be labeled by $2n-1$ independent quantities, which are conserved by the asymptotic dynamics. For a detailed explanation of this issue and general information on the appropriate choice of asymptotic labels, we refer the interested reader to section 2.1 of Ref. \cite{jung97}.

In the present case, a convenient choice of the 5 asymptotic labels is as follows: The first 2 labels are either the initial particle momenta $p_{x,0}$ and $p_{z,0}$, or equivalently the initial kinetic energy $E_0$ and the incident angle $\theta_0$ between the $z$ axis and the initial vector momentum $\vec{p_0}$. A third label consists of the impact parameter $b$ as defined before in section 2, the fourth label is a relative phase shift $\phi$ between the particle motion and the oscillator.  The relative phase shift $\phi$ is defined as $  \phi_0 = \psi_i - ( t_s - t_i ) \omega  $ where $ \psi_i $ is the phase of the oscillator at time $ t_i $ when the particle starts at the point $ (x_i, z_i) $. The quantities $t_s$ and $t_i$ are as defined before in section 2.  

As the fifth and last asymptotic label one can use either the initial oscillator action $I$, or the total system energy, i.e., including that of the oscillator. However, this fifth quantity is irrelevant since the action $I$ never enters any significant quantity; therefore, is not necessary to consider the value of $I$.

In the following, $E_0$ will denote the value of $H_0$, i.e., the particle energy. Its value in the asymptotic region is the kinetic energy of the particle, and its value along an incoming asymptote is denoted $E_0$, and its value along an outgoing asymptote is denoted $E_f$.

\subsection{The scattering function and its Jacobian determinant}
%
\begin{figure}
\begin{center}
\includegraphics[scale=0.7]{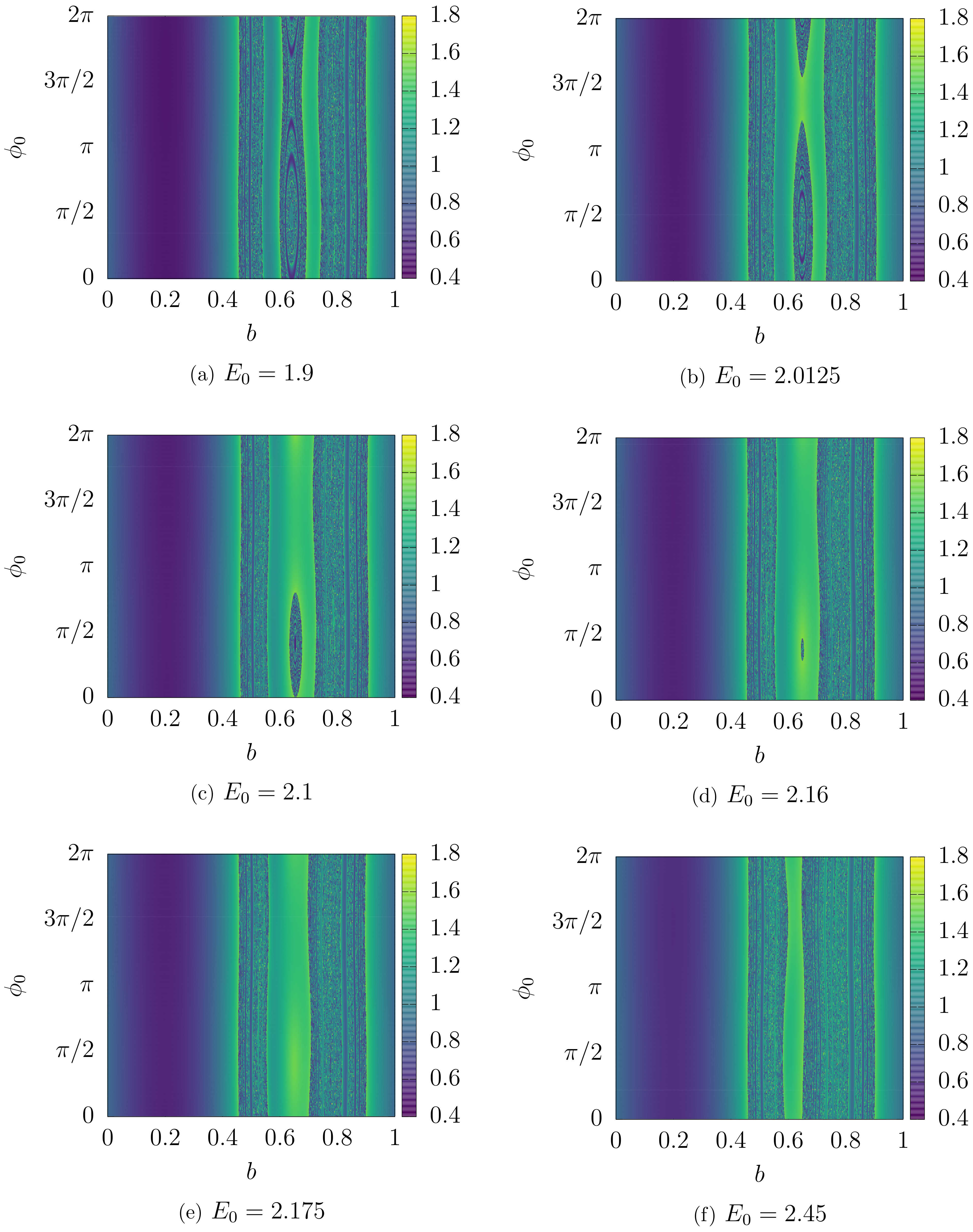}
\caption{Scattering function $\theta_f(b,\phi_0)$ for the initial conditions $\theta_0=80^{\circ}$ and the 6 values of the incident energy, $E_0=$1.9, 2.0125, 2.1, 2.16, 2.175, 2.45 in the parts (a), (b), (c), (d), (e) and (f) respectively. The structure of the fractal set of singularities changes when the value of $E_0$ is increased. In comparison with the unperturbed case of \ref{S_F_2DF}, where the scattering functions are independent of $\phi_0$, here, this symmetry is broken by the perturbation, and new structures with different topology appear, namely annular regions of continuity.
 \label{S_F_I}}
\end{center}
\end{figure}


\begin{figure}
\begin{center}
\includegraphics[scale=0.7]{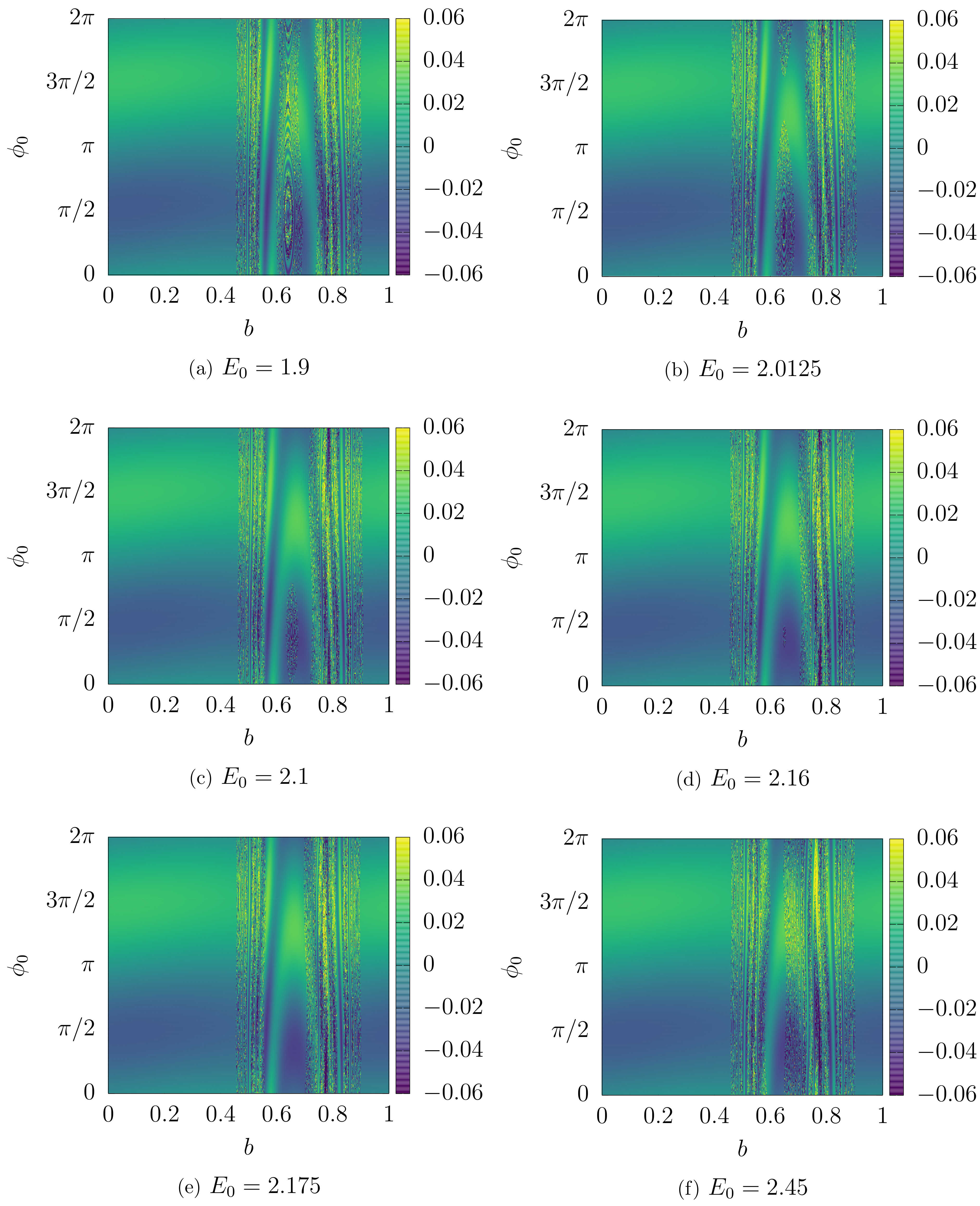}
\caption{Scattering function $\Delta E(b,\phi_0)$ for the initial conditions $\theta_0=80^{\circ}$ and the 6 values of the incident energy, $E_0$=1.9, 2.0125, 2.1, 2.16, 2.175, 2.45 in the parts (a), (b), (c), (d), (e) and (f) respectively. 
 \label{S_F_I2}}
\end{center}
\end{figure}

The scattering function in the 3-dof case is again a map from the set of all possible initial asymptotes to some magnitudes characterizing the final asymptotes. In the present case, the domain of this map is a 5-dimensional set. However, the initial action $I$ is irrelevant, and therefore it will not be considered. Moreover, also in the 3-dof case, the particles in the incoming beam have initial vector momentum $\vec{p}_i$, or equivalently $E_0$ and $\theta_0$, and the quantities $b$ and $\phi_0$ have a random distribution with constant density. The most interesting quantities to measure along the outgoing asymptotes are the final energy $E_f$ of the particle and the trajectory inclination $\theta_f$. Equivalently, it is possible to the final vector momentum $\vec{p}_f$.

For the numerical calculations, let us consider a fixed value of the surface oscillation frequency $\omega = 0.05$ $ \omega_D$, where $\omega_D$ is Debye frecuency, and proceed similarly to the 2-dof system. As long as the oscillation amplitude $B_z=0$, the initial energy of the particle $E_0$ is equal to its final value of $E_f$. However, as soon as $B_z \neq 0$ (for the following numerical examples, the value of the oscillation amplitude is fixed at $B_z=0.001$), the energy $E_f$ is different from $E_0$ in general, and the scattering process becomes inelastic. A convenient quantity to characterize final asymptotes is the energy transfer $\Delta E = E_f - E_0$. For the present system, a meaningful set of 2 scattering functions are: $\Delta E(b,\phi_0)$ and $\theta_f(b,\phi_0)$, both depending on $b$ and $\phi_0$ while keeping $E_0$ and $\theta_0$ fixed, like the beam already described above. 

Some numerical examples, corresponding to the cases  $E_0 = 1.9$, $2.0125$, $2.1$, $2.16$, $2.175$ and $2.45$ are shown in Figs.~\ref{S_F_I} and \ref{S_F_I2}, where Fig.~\ref{S_F_I} shows $\theta_f$ and Fig.~\ref{S_F_I2} shows $\Delta E$. The incoming angle is always kept fixed at $\theta_0 = 80^{\circ}$. As can be seen, the structure of the set of singularities of the scattering functions changes as the value of the initial energy $E_0$ is increased, and the sequence of $E_0$ values used in the figures demonstrates well the fundamental pattern of these changes. For $E_0=1.9$, two basic structures are seen: First, vertical strips along the $\phi$ direction, which are qualitatively similar to the strips that are observed in the uncoupled 2-dof case, i.e., for $B_z=0$. And second, disks around a central point. The complement of the set of singularities is the set of regions of continuity of the scattering functions. If the value of $E_0$ is increased, the structure of strips and disks changes. This change reflects the changes in the structure of the associated tangle between the stable and unstable manifolds of the 3-dof system.

In the analysis in subsection 3.3 it is necessary to consider another important set in the domain of these scattering functions, i.e. in the $b$--$\phi$ plane, namely the curves along which the Jacobian determinant of these functions is zero ( for its importance see also \cite{jung97} ). The Jacobian determinant is defined as

\begin{equation}
\det J = \left| \frac{\partial (\theta_f, \Delta E)}  {\partial(b,\phi_0) } \right |  =
\frac{\partial \theta_f} { \partial b } \frac{\partial \Delta E} { \partial \phi_0} -
\frac{\partial \theta_f} { \partial \phi_0} \frac{\partial \Delta E} {\partial b} 
\end{equation}

In order to understand the pattern of the subset of the domain with $\det J = 0$ let us imagine first extremely small values of $B_z$. As can be seen from Fig.2, in this case, the derivative $\partial \theta_f / \partial b$ is of order 0 in $B_z$ and all the other 3 partial derivatives are of order 1. Therefore in this limit, the second product on the right-hand side of Eq. 4 is of second order and it is possible to neglect it, while the first product is of order one and it becomes the important one. In first order it is the product of the following two factors: First, $\partial \theta_f / \partial b$ taken from Fig.2 which has a single minimum in each interval of continuity. This leads to a single line of $\det J = 0 $ running in $\phi $ direction in each region of continuity. In the following, let us call such lines the vertical $\det J = 0$ lines. And the second factor is the derivative $\partial \Delta E / \partial \phi_0 $. The functional form of the perturbation as given in Eq.3 makes it understandable that in the limit of extremely small $B_z$ the dependency of $\Delta E $ on $\phi_0$ goes like $\sin{\phi_0}$. Then $\Delta E$ has a relative extrema at $\phi_0 = \pi/2$ and $\phi_0 = 3 \pi /2$. In the following, let us call such lines the horizontal $\det J = 0$ lines. 

Under increasing $B_z$ the vertical $\det J =0$ lines are relatively robust as long as the regions of continuity remain stripes running around in $\phi$ direction. And these lines also remain lines running around in $\phi$ direction. The horizontal $\det J =0$ lines are more sensitive to perturbations of the potential. It depends on the width in $b$ direction of the regions of continuity at which value of $B_z$ the horizontal $\det J = 0$ lines deform strongly. The chosen value $B_z = 0.001$ is the appropriate one to study the deformation of the $\det J =0$ lines in all details for the region $R$. This is an additional reason for this
choice of the numerical value of $B_z$ in the present study.

\begin{figure}
\begin{center}
\includegraphics[scale=0.7]{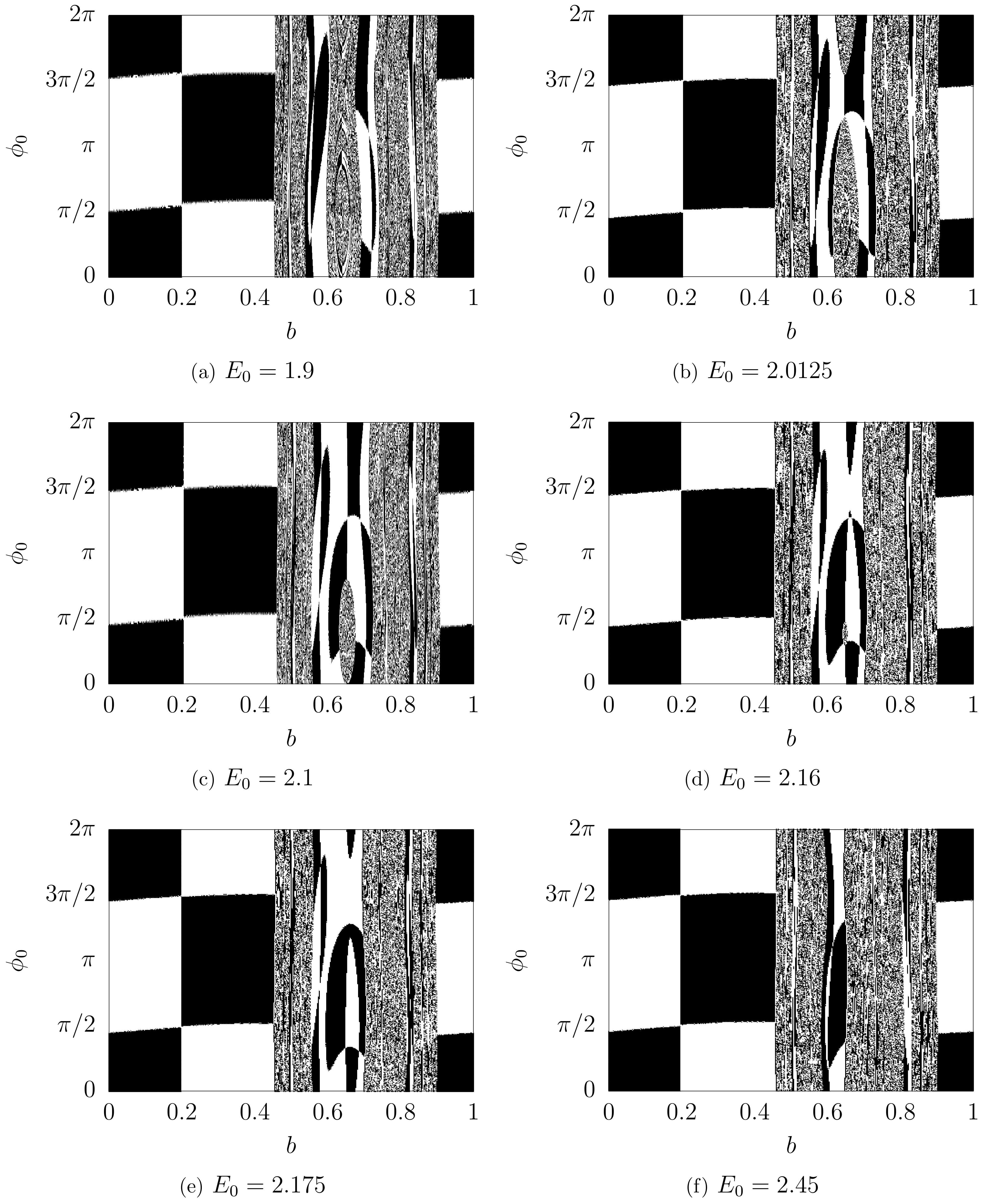}
\caption{Signature of $\det J$ for the scattering functions. The black regions have $\det J > 0$ , 
the white regions have  $\det J < 0$. 
\label{det0}}
\end{center}
\end{figure}

\begin{figure}
\begin{center}
\includegraphics[scale=0.7]{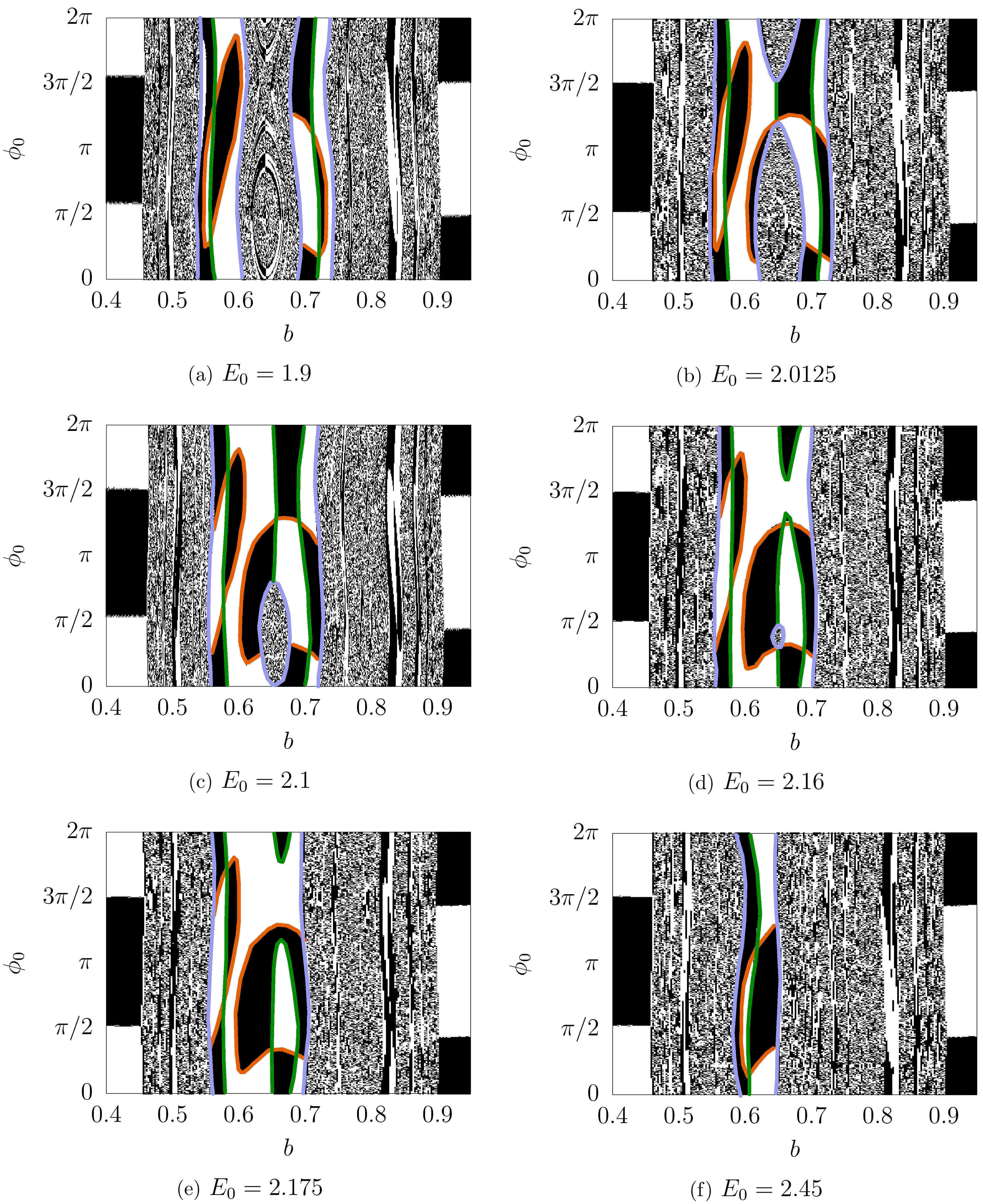}
\caption{Zoom of the signature of $\det J$ for the scattering functions in the interval $ b \in [ 0.4, 0.95] $.
 The black regions have $\det J > 0$ , the white regions have  $\det J < 0$. 
 The green lines correspond to vertical $\det J = 0$ lines in 
the region of continuity $R$. The orange lines are the deformation of the horizontal lines 
$\det J = 0$ in the region $R$. The violet lines are the boundaries of the region $R$.
\label{det0_zoom}}
\end{center}
\end{figure}

The Fig.~\ref{det0} shows the numerical example for the dependence of the $\det J=0$ curves on $E_0$. The 6 parts (a), (b), (c), (d), (e), (f) of the figure correspond to the energy values also used in Figs.~\ref{S_F_I}  and \ref{S_F_I2}. Regions with $\det J >0$ are colored black while regions with $\det J <0$ are shown white. The curves $\det J =0$ are the boundary curves between white and black regions. Remember that $b$ and $\phi_0$ are periodic variables and that, therefore, opposite boundaries of the figure should be identified to turn the domain into a torus. Fig.~\ref{det0_zoom} is a magnification of the most interesting region in Fig.~\ref{det0}. We are mainly interested in the region $R$ and therefore curves in $R$  are marked by colors. The $\det J = 0$ curves developing out of the curves $ \partial \theta / \partial b = 0 $ of the unperturbed case ( i.e. the original vertical curves ) are marked green. The $\det J = 0 $ curves developing out of the curves $ \partial \Delta E / \partial \phi = 0 $ in the limit $ B_z \to 0 $ ( i.e. the deformations of the original horizontal curves ) are marked orange. Also, the boundaries of the region $R$ are marked violet. First, observe the robustness of the vertical $\det J =0$ line but also the robustness of the two horizontal $\det J=0$ lines in the largest region of continuity in the Fig.~\ref{det0}. In this largest region of continuity, there are no interesting changes in the topology of the region itself or in the $\det J=0$ lines for moderate values of $ B_z $. 

In order to show the typical scenario for a change of the topology of regions of continuity and related changes of the $\det J=0$ lines, let us focus our attention on one region of continuity, namely the one which is colored yellow in Fig.~\ref{F_S_S}, and which has already been called the region $R$. In Fig.~\ref{F_S_S}, it is the largest region of continuity besides the infinite outer region of continuity. At the lowest energy $E_0=1.9$ displayed in the figures, the region $R$ consists of two components. Compare them with an intersection between a horizontal line at $E_0=1.9$ in Fig.~\ref{F_S_S} and the yellow region in this figure. Of course, because of the nonzero perturbation, the two components wiggle a little as a function of $\phi$, but they still have the same topology as in the unperturbed case, compare part (b) of Fig.~\ref{S_F_2DF}. Each one of the 2 components still has its vertical $\det J=0$ curve (colored green), which still has the same topology as in the unperturbed case.

We also observe the relatively large sensitivity of the horizontal $\det J=0$ curves against perturbations. Only in the largest outer region of continuity, the horizontal curves are almost the same as in the limit of extremely small $B_z$, as explained above. They are two almost horizontal lines near $\phi = \pm \pi / 2 $. In contrast, in all the smaller regions of continuity, the horizontal lines are deformed strongly and have changed their topology such that they can no longer be recognized easily as curves having developed from horizontal lines. This behavior illustrates how the onset of strong deformations of the horizontal $\det J =0$ lines depends on the size of the region of continuity; smaller regions are more sensitive to perturbations. We also see that the value $B_z=0.001$ is the appropriate one to see the essential behavior of the $\det J=0$ lines in the region $R$.

Now let us describe in more detail the transformation of the $\det J = 0$ curves within $R$ under a variation of the energy $E_0$. When the value of $E_0$ is increased then the gap separating the two components of $R$ becomes smaller, and at a critical value of $E_0$, the gap breaks near $\phi_0 = 3 \pi / 2$. At this moment, the gap changes its topology from a stripe running around $\phi_0$ direction to a disc contractible to a point. Accordingly, also $R$ changes its topology from two disconnected stripes running around in $\phi_0$ direction to a connected set. At the same time, a new vertical $\det J =0$ curve is created running between the two extremal points in $\phi_0$ of the disc-shaped gap. See part (b) of the figure. With $E_0$ increasing further, the gap shrinks, and the right vertical $\det J=0$ curve and the new middle $\det J=0$ curve come closer, see part (c) of the figure. At another critical value of $E_0$, these two vertical $\det J=0$ curves touch at $\phi_0= 3 \pi /2$, and change their topology as shown in part (d) of the figure. They change from two curves running around in $\phi_0$ direction to a single contractible loop. Next, the gap disappears completely, as can be seen in part (e) of the figure. Simultaneously with increasing $E_0$, the contractible green $\det J =0$ curve shrinks and finally disappears, while also the right orange curve shrinks and disappears. The result is shown in part (f) of the figure. In the end, there is a single region of continuity running around in $\phi_0$ direction and containing a single vertical $\det J=0$ curve also running around in $\phi_0$ direction. Thereby the scenario of the fusion of two typical regions of continuity into a single one is finished. This scenario is the direct generalization of the 2-dof scenario of the fusion of intervals of continuity, as described in Ref.\cite{jung}. For other regions of continuity similar to transformations and fusions happen under a change of the energy.

\subsection{Stable and unstable manifolds in the 3-dof problem} 

Let us discuss the construction of the tangle between the stable and unstable manifolds in the 4-D Poincar\'e map for the 3-dof case following the ideas developed in Refs.~\cite{zapfe,zapfe2} for systems with one open and 2 closed dof. 

In order to explain the construction, let us start with the case of zero oscillation amplitude, i.e., $B_z = 0$. In this case, the particle energy $E_0$ is conserved, and each slice, $E_0$ = constant, in the 4-D Poincar\'e map is invariant. The tangle in each one of such slices was already discussed in section \ref{sec:2dof} (see Fig.~\ref{H}). Now, the higher dimensional tangle for the 3-dof system can be obtained in a two-step process. First, we form a stack with all curves taken from the whole continuum of 2-dof tangles, where the value of $E_0$ is the stack parameter. This union creates a 3-D object.

\begin{figure}
\begin{center}
\includegraphics[scale=1]{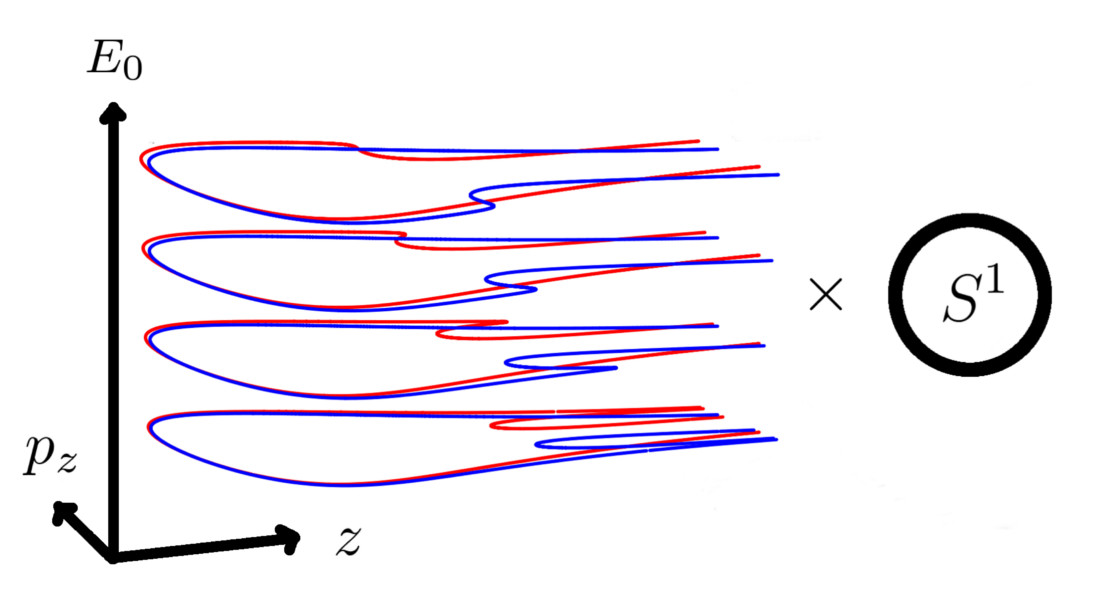}
\caption{ The Cartesian product of the pile of 2-D tangles with the energy $E_0$ as stack 
parameter and a circle representing the angle $\psi$ forms the tangle between the stable 
and unstable manifolds for the 3-dof system in the Poincar\'e map.} 
\label{T}
\end{center}
\end{figure}

Second, we add to this object the phase variable of the oscillator by forming a Cartesian product of the 3-D stack with a circle representing the still missing variable $\psi$. In this way, we end up with a tangled structure in the 4-D domain of the Poincar\'e map for the 3-dof system; see Fig.~\ref{T}. 
Notice, that this is still only for zero oscillator amplitude, $B_z=0$. 

In the 2-dof case and the corresponding 2-dimensional map of section \ref{sec:2dof}, the tangle was created, starting from the fixed point at infinity ($z=\infty$, $p_z=0$), by plotting the stable and unstable manifolds associated with this point at infinity. In the stack of the 2-dimensional maps, the stack of fixed points at infinity is a vertical line, and the Cartesian product of this line with the circle corresponding to the oscillator phase generates an invariant 2-dimensional cylinder at infinity. It is a 2-dimensional invariant subset of the 4-dimensional domain of the map for the 3-dof system.
 
Thereby the object growing out of the outer fixed point of the 2-dof tangle is this cylindrical invariant 2-dimensional surface. If the fixed point at the infinity of the 2-dof system were hyperbolic, then the resulting 2-D surface in the 4-D domain would be a normally hyperbolic invariant manifold (NHIM). In our case, this object is formally not a NHIM, but it plays an analogous role. For general information on NHIMs and their role in dynamical systems, see for example \cite{wiggins,wiggins3}.

The stable and unstable manifolds $ \mathcal{W}^{s/u}$ of the invariant surface at infinity are also obtained by the same stack-and-product process as
%
\begin{equation}
  \mathcal{W}^{s/u} = \bigcup_{E_0} W^{s/u}_{E_0} \times S^1.
 \label{eq:4}
\end{equation}

Let us now check the dimension of these manifolds. The dimension of the stable and unstable manifolds in the 2-dof case for fixed energy is 2 in the flow or 1 in the map. The pile of the stable and unstable manifolds, parametrized by the initial particle energy $E_0$, is a surface with dimension 3 in the flow or dimension 2 in the map. Then forming the product with a 1-D circle results in a 4-D surface in the flow of the 3-dof system or a 3-D surface in the domain of the map. Most important: The codimension of the stable and unstable manifolds is 1 in any case, and then the stable and unstable manifolds create a partition in the constant total energy manifold of the 3-dof system, and also in the domain of the Poincar\'e map. The invariant manifolds $ \mathcal{W}^{s/u}$ are impenetrable hypersurfaces that direct the flow similar to what happens in the 2-dof systems. Moreover, the stable/unstable manifold is the union of the trajectories whose $z$ component of the momentum converges to zero for $z \to \infty$.

So far, we have described the construction of the tangle for the 3-dof problem in the case of no coupling, i.e., for $B_z = 0$. What happens when the coupling is switched on? If the system would has a usual NHIM, then we could quote the theorem of Fenichel on the persistence of NHIMs and their stable and unstable manifolds, see, for example, Refs.~\cite{fenichel  ,wiggins,eldering}. However, in the present case, the invariant surface at infinity is stacked up by linearly parabolic points, which are only nonlinearly unstable. On the other hand, in this case, the interaction (see Eq.~(\ref{eq:2})) goes to zero exponentially for large values of $z$. Therefore, the invariant surface itself at infinity is not affected by the interaction. The tangle of the stable and unstable manifolds coming from the points at infinity contains hyperbolic components ( transverse intersections between stable and unstable manifolds ) and in general, also nonhyperbolic components ( tangencies or situations close to tangencies )  as it is generally found in usual incomplete horseshoe constructions. The general experience indicates that for scattering processes the hyperbolic part of the chaotic saddle dominates. The nonhyperbolic parts are less important and are influencing only very high levels of hierarchical in the resulting fractal structure. For several examples see the references\cite{Alt_1996,Motter_2001,Gaspard_1995,Kokshenev_2000,Dettmann_2009,Altmann_2013,Jung_1993}  . The hyperbolic parts are robust under small perturbations of the system, while the nonhyperbolic parts may change already under the smallest perturbations.

In total, we can expect that the stack of tangles maintains its large-scale structure also under moderate perturbations of the system, and therefore also under moderate coupling strengths between the particle dofs and the oscillator dof. As a result, the intersection of the perturbed stack with a plane surface should be similar to the intersection of the unperturbed stack with a curved surface. In this sense, the stack construction method provides us with a valid idea of the higher dimensional tangle and is also able to explain the dynamics of the system with coupling.

Now, we understand from another point of view the changes in the structure of the set of singularities in the scattering functions for the coupled 3-dof system in Figs.~\ref{S_F_I}, \ref{S_F_I2}, \ref{det0}. When the 3-dof system contains a perturbation, then the symmetry in the angle $\phi$ is broken, and the manifold $ \mathcal{W}^{s}$ is deformed and loses its symmetry with respect to $\phi$. And its intersection with the set of initial conditions change. If we change the set of initial conditions in the phase space, then the pattern of intersections is different. The plots of the various parts in Figs.~\ref{S_F_I}, \ref{S_F_I2}, \ref{det0} are scattering functions for different sets of initial conditions (different $E_0$), and their singularities show the pattern of the intersection of the stable manifold.  
 
With the help of Fig.~\ref{F_S_S} and our new understanding, we can give still another interpretation of the change of the regions of continuity of the scattering functions under perturbations. Depending on the value of $\phi_0$, the value of the particle energy $E_0$ changes. Thereby the particle can change in Fig.~\ref{F_S_S} from the initial value of $E_0$ to a modified value of $E_0$, and this modification depends on $\phi_0$. Imagine that the initial value lies around2, e.g., at the value 2.0125 used in parts (b) of the Figs.~\ref{S_F_I}, \ref{S_F_I2}, \ref{det0}. Here the region $R$ is very close to the value where it switches in Fig.~\ref{F_S_S} from having 2 components to having 1 component. For $\phi_0$ values around $-\pi/2$ the particle energy is increased ( see Fig.~\ref{S_F_I} (b) ) and correspondingly the system runs into the situation where along the $b$ direction $R$ has 1 component and for $\phi_0$ values around $\pi/2$ the particle energy is decreased ( see again Fig.~\ref{S_F_I} (b) ) and the system runs into the situation where along the $b$ direction $R$ has 2 components. This is exactly what we observe in the parts (b) of the Figs.~\ref{S_F_I}, \ref{S_F_I2}, \ref{det0}. Considerations of this type only hold for weak perturbations where the homoclinic/heteroclinic tangle still has the stack structure.

\subsection{The cross-section and its connection with the scattering functions}

Let us assume an incoming beam, as explained before. The detector should measure the distribution of the values of the outgoing particle energy $E_f$, or equivalently and even better, the energy transfer $\Delta E$ and the outgoing angle of inclination $\theta_f$. This detector registers neither the outgoing phase shift $\phi_f$ nor the value of the outgoing impact parameter $b_f$. The result in this type of measurement is the doubly differential cross-section $ \frac {d \sigma} { d \theta d \Delta E} (\theta_f, \Delta E) $ for fixed values of the initial energy $E_0$ and fixed initial angle of incidence $\theta_0$.  

Now, let us discuss the geometrical connection between the scattering function and the cross-section. Remember that the relevant scattering function for the construction of the cross-section is the function whose domain and range are the $b$--$\phi_0$ and $\Delta E$--$\theta_f$ planes, respectively. This function can be viewed as a graph in the Cartesian product of the 2-D domain with the 2-D range, i.e., in the 4-D $(b, \phi_0, \Delta E, \theta_f)$ space. The incoming beam represents a constant density on the domain, and the function maps this density into the range. The differential cross-section defined above is the resulting density in the range. This connection is expressed as follows
\begin{equation}
\frac{d \sigma}{d \theta \; d \Delta E} (\theta_f, \Delta E) = 
\sum_i \left| \det \frac{\partial(\theta_f,\Delta E)}{\partial(b,\phi_0)}\right|^{-1},
  \label{eq:5}
\end{equation}
where the sum runs over all preimage points in the domain, i.e., in the $b$--$\phi_0$ plane, leading to the values $\theta_f, \Delta E$ in the range of the scattering function. It is clear that the cross-section has singularities where the projection of the graph into the range is singular, i.e., where the determinant of the Jacobian matrix of the scattering function appearing in the previous equation is zero, i.e., they are the lines already studied in subsection 3.1 and Fig.~\ref{det0}. The resulting singularities in the cross-section are the well-known rainbow singularities of the differential cross-section, and they are the lines over which the number of preimages changes (in general by 2). In a general 3-dof system, there are lines along which the rank of the Jacobian matrix drops by 1, and there can also be points at which the rank drops by 2. For a good explanation of rainbow singularities, see chapter 5 in Ref. \cite{newton}. Note that generic rainbow singularities are of one over square root type, and therefore the integral over them is finite. At the points where the rank of the Jacobian matrix in Eq. (\ref{eq:5}) drops by 2, there might be singularities of another functional form, however also here the integral over the differential cross-section is always finite. There is no violation of flux conservation.
 
In the present system, most regions of continuity of the scattering function are stripes running around in $\phi_0$ direction or annular shaped regions, see Fig.~\ref{S_F_I}. However, under a change of the energy, the processes of fusion of regions of continuity happen as explained in all details in subsection 3.1. In the rest of this subsection, we will study how the process of fusion shown in the Figs.~\ref{S_F_I}, \ref{S_F_I2}, \ref{det0} shows up in the cross-section. 

If we project the graph of the scattering function of one vertical stripe or also of one disc-shaped region of continuity into the plane $\Delta E$--$\theta_f$ we can see a caustic like the one coming from the projection of a deformed semi-torus, this structure is characteristic of 3-dof systems with 1 open dof and 2 closed dofs \cite{zapfe2}, it is one normal form for these projection caustics. A more detailed explanation for this structure is the following. 

First, let us consider the unperturbed case. In the inner part of a vertical stripe of continuity, the scattering function is constant in the $\phi$ direction, and we have a kind of half semi-torus in the plot of the scattering function, remember that $\phi$ is a periodic variable and that the vertical stripe closes to a ring. The projection of this graph on the plane $\Delta E$--$\theta_f$ is a quadrilateral. And it is a 4:1 projection. If the system is perturbed, the rotational symmetry is lost, and the plot of the scattering function is deformed. The semi-torus is then deformed, and it is no longer of constant height; accordingly, the caustics change, and also, some regions are formed where the projection is 2:1. For the perturbed case, the caustics have the same qualitative structure for different values of the perturbation parameter as long as the system is in the regime of weak perturbation. Analogous considerations hold for the contractible ring-shaped regions of continuity. 

The complete caustic structure of the entire cross-section is a superposition of the various basic caustic structures coming from all the different regions of continuity. Of course, the various structures are shifted in their exact position, and they have different heights and different total strength. This total strength is proportional to the area of the corresponding interval of continuity because it must be proportional to the incoming flux falling into this particular interval of continuity.

\begin{figure}
\begin{center}

\includegraphics[scale=0.7]{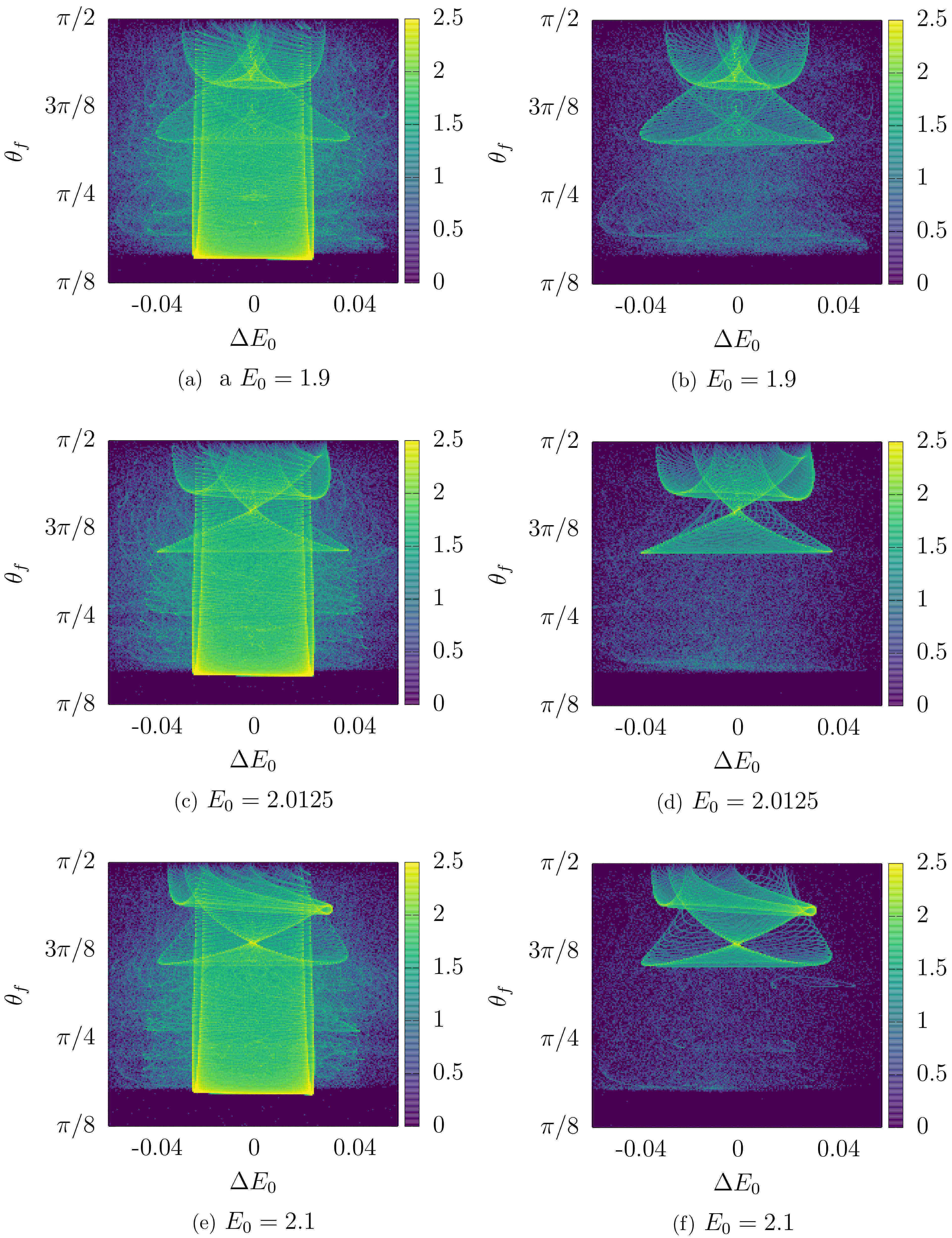}

\caption{Scattering cross-section for $\theta_0 = 80^{\circ}$ and for the 6values $E_0 = $ 1.9, 2.0125, 2.1, 2.16, 2.175 and 2.45 in the parts (a) and (b), (c) and (d),(e) and (f), (g) and (h), (i) and (j), (k) and (l) respectively. The parts (a), (c),(e), (g), (i) and (k) in the left column show the cross section for a constant illumination of the whole $b$--$\phi_0$ plane. The parts (b), (d), (f), (h), (j), (l) show it for partial illumination of a neighbourhood of the region $R$ only.\label{secciones}}
\end{center}
\end{figure}

\renewcommand\thefigure{\arabic{figure}}
\setcounter{figure}{9}

\begin{figure}
\begin{center}

\includegraphics[scale=0.7]{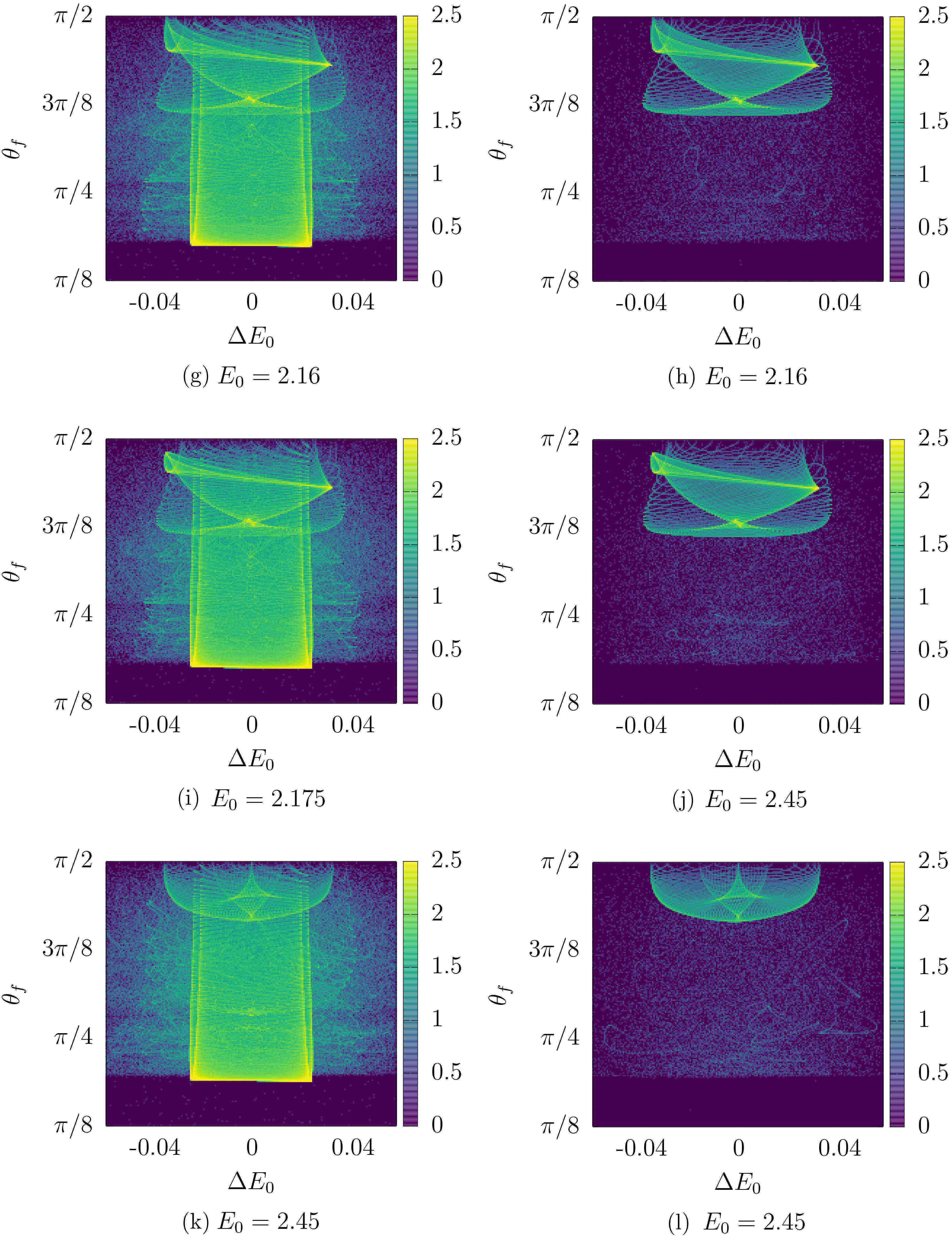}

\caption{Continuation.}

\end{center}
\end{figure}

The Fig.~\ref{secciones} shows a plot of the cross-section for $\theta_0=80^{\circ}$ and for the six different values of the initial energy $E_0$, which also have been shown in Figs.~\ref{S_F_I}, \ref{S_F_I2}, \ref{det0}. In the plots, the basic structure of the singularities of the projection of the graphs of the regions of continuity of the scattering function is apparent. 

First, let us look at part (a) of the Figs.~\ref{S_F_I}, \ref{S_F_I2}, \ref{det0} for $E_0$ = 1.9. Here the scattering function has 3 important regions of continuity. The first one, the outer region, which occupies the largest area in the domain and which accordingly causes the strongest structure in the cross-section. It is the strong structure reaching up to rather small values of $\theta_f$. It comes close to a rectangle, which is the result of a 4:1 projection of a semi torus. In addition, we also see strong structures coming from the two separate components of the region $R$ defined in subsection 3.1. Because we are mainly interested in the contributions from region $R$, we repeat in part (b) the cross-section again for $E_0$ = 1.9 where, however, we have illuminated a neighborhood of the region $R$ only. Thereby the structures caused by region $R$ are clearly visible. They are two of the above mentioned deformed torus-shaped structures. The structure reaching up to smaller values of $\theta_f$ is the one coming from the left part of $R$ in Figs.~\ref{S_F_I}, \ref{S_F_I2}, \ref{det0}.   

Now let us proceed to $E_0$ = 2.0125. The cross-sections with complete and partial illumination of the domain of initial conditions is plotted in parts (c) and (d) of Fig.~\ref{secciones} respectively. When we compare parts (b) and (d) of the figure, then we see the beginning of the fusion process in the upper right part of the main structure. Parts (e) and (f) show the complete and the partial cross-section for $E_0$ = 2.1. Here the fusion has proceeded to large deviations compared to $E_0$ = 1.9. Parts (g) and (h) show the cross-sections for $E_0$ = 2.16 and parts (i) and (j) show it for $E_0$ = 2.175. Here the region $R$ has already turned into a single vertical stripe. However, as Fig.~\ref{det0_zoom} (e) shows there are still remanets of the additional $\det J =0$ curves which create the additional rainbow structures around $\theta_f \approx 7 \pi / 16$. Finally, the parts (k) and (l) show the cross-section for $E_0$ = 2.45, where the transformation process of the region $R$ is finished. Here the rainbow structure coming from $R$ is again just the projection of half a torus. Compare it with one of the two strong structures ( the one sitting mainly at higher values of $\theta_f$ ) seen in part (b) of the figure. We see a similarity, which indicates that the completed fusion process of two generic regions of continuity results in a single generic region of continuity.

The sequence of changes presented is the typical one for any fusion of regions of continuity of the scattering functions. It is the 3-dof generalization of the 2-dof fusion events explained in Ref.\cite{jung}. 

\section{Remarks and conclusions}
  \label{sec:remarks}
  
We studied an example of chaotic scattering in a 3-dof Hamiltonian model with one open and 2 closed dofs, namely the scattering of an atom from a vibrating corrugated surface. The system can be considered a small perturbation of a partially integrable system and therefore a convenient starting point is this partially integrable 3-dof system which can be treated as a stack of 2-dof systems, where the stack parameter is the particle energy $E_0$, which is a conserved quantity of the 2-dof system.
 
A cental object in this study is the scattering function $(\Delta E_0(b,\phi_0), \theta_f(b, \phi_0))$, and its set of singularities. The set of singularities reflects the structure of the stable and unstable manifolds that divide the constant energy manifold of the total system and direct the dynamics. The doubly differential cross-section is the projection of the graph of the scattering function on its range and the singularities of this projection are the caustics. 

The structures of the caustics are related to the regions of continuity of the scattering function and to the curves where the Jacobian determinant of the scattering function is zero, $\det J = 0$. The regions of continuity generate characteristic types of structures on the cross-section; it is like a projection of a semi torus on the $\Delta E_0$--$\theta_f$ plane. An increment in the value of the initial particle energy creates changes in the regions of continuity of the scattering functions and in the set of curves $\det J = 0$, and we can see clearly how those changes are reflected in the fusion of the caustics on the cross-section.

The important progress of the present work in comparison with previous investigations of 3-dof chaotic scattering is a detailed description of the fusion between two regions of continuity of the scattering function. We have divided the fusion process of two generic regions of continuity into a single generic region of continuity into its 5 important steps. The process is presented by a sequence of six plots illustrating the 5 important steps in the transformations. Also, we have presented the corresponding steps in the changes observed in the doubly differential cross-section.

The present system helps to understand the elementary transitions in the scattering functions and the corresponding transitions in the cross-section also in other systems with qualitatively similar changes in the scattering function, for example, 3-dof systems with a perturbation of partial integrability \cite{zapfe}, \cite{zapfe2},\cite{Drotos_2012}, a perturbed magnetic dipole \cite{gonzalez}, and a realistic molecular system treated in \cite{Yi_2013}. The considerations of subsections 3.1 and 3.3 hold for all the above mentioned system in an analogous form. Note that Ref. \cite{Yi_2013} is a system with 2 open dof, and accordingly, the structure of the vertical $\det J =0$ lines is more complicated. The scattering function oscillates in the interior of regions of continuity, and this leads to a sequence of vertical $\det J =0$ lines in each region of continuity. Otherwise, the basic phenomena are the same.

For any scattering system, there is always the interesting question of the inverse scattering problem. For chaotic scattering, one considers this problem as the problem to reconstruct information on the chaotic invariant set from scattering data, in particular from cross-section data. So far, for 3-dof systems, there is a general strategy to do this job only for the uniformly hyperbolic case, see Ref.\cite{Drotos_2016}. However, these hyperbolic cases are always far away from any stack construction; they are a kind of opposite extremal case to the partially integrable case. Therefore they do not provide any clue for our present case, which is a case of mixed-phase space close to partial integrability.

Finally, we offer some remarks about the implications of the classical scattering results for the quantum scattering problem in the semiclassical regime. Classically the cross-section is a sum over contributions from all preimages, see Eq.6. Semiclassically, the scattering amplitude is a corresponding sum over the square roots of the classical contributions where each contribution gets, also, a phase which is the complex exponential function of the reduced action along the corresponding path. The semiclassical cross-section is the absolute square of the semiclassical scattering amplitude, and thereby it is a double sum. Accordingly, the semiclassical cross-section is a sum of the classical cross-section coming from the diagonal terms of the double sum and a double sum of the nondiagonal interference terms. Thereby the resulting cross-section contains interference oscillations superimposed over the classical cross-section.

In addition, a semiclassical approach should uniformize the rainbow singularities and remove, thereby, the infinities. In a generic rainbow line, 2 classical contributing paths coincide and disappear. Then the uniformized semiclassical contribution of these 2 classical paths to the semiclassical scattering amplitude can be modeled by an Airy function, which is the normal form of a wave rainbow contribution. The contribution of a point where the rank of the Jacobian determinant of the scattering function drops by 2 should be described by some other appropriate catastrophe function. For more detailed information on the semiclassical treatment of chaotic scattering see Refs.\cite{Jung_1990_2}, \cite{Jung_1990},\cite{Jense_1994},\cite{Jense_1995}.

\section{Acknowledgments}

F G acknowledges the support of CONACYT program for Postdoctoral Fellowship. C J acknowledges financial support by DGAPA under Grant No. IG100819. F B acknowledges the financial support from Spanish Ministry of Science, Innovation and Universities, Gobierno de Espa\~na, under Contracts No.\ PGC2018-093854-BI00, the Ministerio de Econom\'{i}a y Competitividad (Spain) under Contracts No.~MTM2015-63914-P and ICMAT Severo Ochoa Contract No.~SEV-2015-0554, and from the People Programme (Marie Curie Actions) of the European Union's Horizon 2020 research and innovation program under Grant No. 734557.

\bibliographystyle{unsrt}
\bibliography{bibliography_d}

\end{document}